\documentclass[superscriptaddress,twocolumn,a4paper,showpacs,eqsecnum,pra]{revtex4}
\usepackage{xspace,amsmath,amsfonts,amsthm,amssymb,graphicx}
\usepackage{ulem,ifthen,color}
\newcommand{\beq}{\begin{equation}}
\newcommand{\eeq}{\end{equation}}
\newcommand{\beqa}{\begin{eqnarray}}
\newcommand{\eeqan}{\end{eqnarray*}}
\newcommand{\beqan}{\begin{eqnarray*}}
\newcommand{\eeqa}{\end{eqnarray}}
\newcommand{\bra}[1]{\langle{#1}|}

\newcommand{\ket}[1]{|{#1}\rangle}

\newcommand{\ip}[1]{\langle{#1}\rangle}

\newcommand{\non}{\nonumber}

\newcommand{\eqr}[1]{Eq.~(\ref{#1})}

\newtheorem{theo}{Theorem}

\newcommand{\bqa}{\begin{eqnarray}}
\newcommand{\eqa}{\end{eqnarray}}
\newcommand{\nn}{\nonumber}

\newcommand{\erf}[1]{Eq.~(\ref{#1})}

\newcommand{\dg}{^{\dagger}}

\newcommand{\smallfrac}[2]{\mbox{$\frac{#1}{#2}$}}
\newcommand{\half}{\smallfrac{1}{2}}

\newcommand{\sq}[1]{\left[ {#1} \right]}

\newcommand{\ro}[1]{\left( {#1} \right)}

\newcommand{\tr}[1]{{\rm Tr}\sq{ {#1} }}

\newcommand{\Syn}{\Upsilon}

\newcommand{\gu}{\affiliation{Centre for Quantum Dynamics, Griffith University,  Brisbane, 4111 Australia}}
\newcommand{\uh}{\affiliation{Quantum Physics Group, University of Hertfordshire, Hatfield, AL10 9AB, UK}}
\newcommand{\um}{\affiliation{Department of Physics, University of Massachusetts at Boston, 100 Morrissey Blvd, Boston, MA 02125, USA}}
\newcommand{\vi}{\affiliation{Venetian Institute of Molecular Medicine, Via Giuseppe Orus, 2 35129 Padova, Italy}}

\begin{document}

\title{Tradeoff between extractable mechanical work, accessible
entanglement, and ability to act as a reference system, under arbitrary superselection rules }

\author{J.A. Vaccaro} \gu\uh
\author{F. Anselmi} \vi\uh
\author{H.M. Wiseman} \gu
\author{K. Jacobs}\gu\um
 \date{\today}


\begin{abstract}
Superselection rules (SSRs) limit the mechanical and quantum processing resources represented by
quantum states. However SSRs can be violated using reference systems to break the underlying
symmetry. We show that there is a  duality  between the ability of a system to do mechanical work
and to act as a reference system. Further, for a bipartite system in a globally symmetric pure
state, we find a triality between the system's ability to do local mechanical work, its ability to
do ``logical work'' due to its accessible entanglement, and its ability to act as a shared
reference system.
\end{abstract}

\maketitle

\section{Introduction}
Global conservation laws give rise to superselection rules (SSRs) which forbid the observation of coherences between
particular subspaces of states \cite{WWW,AhaSus}. Such global laws do not apply in subsystems
\cite{AhaSus,KitMayPre}. For example, the angular momentum of an object can be changed provided the total angular
momentum of the object and another system, the ancilla, is conserved. The ancilla here acts as a {\em reference system}
which alleviates the affect of the SSR by locally breaking the associated symmetry \cite{AhaSus}. Conversely, the lack of
a reference system induces the SSR. For example, without a spatial orientation frame, the state of a spin-$\frac{1}{2}$
particle will be completely mixed.

The last few years has witnessed a resurgence of interest in SSRs and quantum reference systems particularly within the
context of quantum information theory.  The recent review by Bartlett, Rudolph and Spekkens \cite{BRS-RMP}
describes the current state of affairs. For example,  Eisert {\it et al.} \cite{EisFelPapPle} and recently Jones {\it et al. }
\cite{JonWisBar} studied the decrease in distillable entanglement due to the loss of relative-ordering information for
sets of ebits. The optimal cost of aligning reference frames has been calculated in a number of different settings
\cite{ARF}. Communication in the presence or absence of shared reference frames has been extensively studied by
Bartlett {\em et al.} \cite{BarRudSpe}. The conservation of particle number was shown by two of us \cite{WisVac} to
limit shared particle entanglement.   The repercussions for various systems including those in condensed matter physics
was explored by Dowling {\em et al.} \cite{DDW}. This constraint on shared entanglement of particles has been
generalized to arbitrary SSRs by Bartlett and Wiseman \cite{BarWis}. For the special case of a U(1)-SSR, a new
resource, the shared phase reference, has been studied by Vaccaro {\em et al.} \cite{VacAnsWis}, and quantified in the
asymptotic \cite{SchVerCir} and nonasymptotic \cite{Enk05} regimes.

In this paper we investigate the effect of a SSR on the resources represented by a quantum state. Following Oppenheim
{\em et al.} we quantify the resources in terms of {\em mechanical} work extractable from a heat bath and {\em logical}
work as performed in quantum information processing (QIP) \cite{OppHor}. We expose a fundamental tradeoff between
the extractable work under the SSR and the ability to act as a reference system for the SSR. We treat both the unipartite
and the bipartite case. The latter shows a triality between the accessible entanglement, locally extractable mechanical
work and the ability to act as a shared reference system. These results are crucial for fully understanding and quantifying
resources used in QIP.

We wish to emphasize from the outset that the resources are determined in the {\em non-asymptotic} regime in the
following sense. While the asymptotic limit $\rho^{\otimes n}$ for $n\to\infty$ is often taken when studying resources
such as entanglement, this limit is not appropriate for the problems addressed here. Indeed, in the asymptotic limit
reference systems such as those for spatial orientation and quantum phase reduce to their less-interesting classical
counterparts. Instead the situation we consider is when the resources such as accessible entanglement, local work and
reference ability are measured for just {\it one copy of the state $\rho$}.  The same situation has been treated in previous
work \cite{WisVac, VacAnsWis} for the specific case of the accessible entanglement of indistinguishable particles. In
operational terms, we imagine that the state of the system is {\it transferred} by operations that are allowed by the SSR to
ancillary systems which are not themselves subject to the SSR. Once transferred to the SSR-free ancillas, the resources
are fungible in the sense that they can be used, processed, transferred etc. in a manner free of the SSR. Our results
quantify the amount of the resources that are transferable in this way from the single copy of the state $\rho$ and made
SSR free. This is what we mean by the terms {\it extractable work} and {\it accessible entanglement}. Thereafter one
could consider the asymptotic limit of the resources contained in the SSR-free ancillary systems and this would justify
the entropic measures for work and entanglement.

\section{Extractable work and asymmetry}
\subsection{Framework for the SSR }

An SSR is associated with a set $\tau=\{T(g)\}$ of unitary operations indexed by $g$ whose action on the system is both
forbidden and undetectable. There are two physically-motivated conditions on the set $\tau$.  If an operator $T(g)$ is
forbidden under the SSR, then so is the time-reversed process which is given by the inverse $T^{-1}(g)$. This means if
$T(g)\in \tau$ then $T^{-1}(g)\in \tau$. If two operations $T(g_1)$ and $T(g_2)$ are forbidden then their product
$T(g_1)T(g_2)$ is necessarily forbidden and correspondingly the product is an element of $\tau$. The set $\tau$ is
therefore closed under multiplication. These conditions endow $\tau$ with a group structure, i.e. the set
\beq
  \tau=\{T(g):g\in G\}
\eeq
is a unitary representation of the abstract group $G=\{g\}$. We shall label the SSR associated with group $G$ as
$G$-SSR  \cite{LieGroupNote}.

Let $\rho$ be an arbitrary density operator representing the (possibly mixed) state of a system. A $G$-SSR restricts not
this state, but rather the allowed {\em operations} on it to those that are $G$-invariant \cite{BarWis}. That is, an
allowed operation ${\cal O}$ must satisfy
\beq
   {\cal O}[T(g) \rho T^\dagger(g)] = T(g) ({\cal O}\rho) T^\dagger(g), \forall g\in G\ .
   \label{G invariant op}
\eeq
Under this restriction, our effective knowledge of the system is represented not by $\rho$ but by the ``twirl'' of $\rho$
\cite{BarWis}
\beq
   {\cal G}_G[\rho] \equiv \frac{1}{|G|}\sum_{g\in G}
         T(g) \rho T^\dagger(g)\ ,
   \label{inv_op}
\eeq
where $|G|$ is the order of the group $G$.

We will require that the representation factorizes as
\beq
   T(g)=T_1(g)\otimes T_2(g)\otimes\cdots
   \label{T_factorisation}
\eeq
for multipartite systems whose corresponding Hilbert space is given by ${\cal H}={\cal H}_1\otimes{\cal
H}_2\otimes\cdots$ where ${\cal H}_n$ is the Hilbert space for the system labeled by $n$.

\subsection{Extractable work}

The purpose of a reference system is to mask the effects of the $G$-SSR by yielding less mixing than given in
\erf{inv_op}. A {\it physically meaningful} definition of the ability of a system to act as a reference system should
therefore be based on a physical quantity that measures a state's mixedness. This measure is conveniently provided by the
amount of  mechanical work that can be extracted from a thermal reservoir at temperature $T$ using quantum state
$\rho$.  This is given by \cite{vonN,OppHor}
\beq
   W(\rho) = k_{\rm B}T[\log D - S(\rho)],
   \label{W}
\eeq
where $D$ is the dimension of the Hilbert space and
\beq
   S(\rho) \equiv - \tr{\rho\log\rho}
\eeq
is the von Neumann entropy of $\rho$.  This expression shows that the more pure the state $\rho$ is, the more work that
can be extracted using it. For convenience, in the following we set $k_{\rm B}T=1$ and use the binary logarithm. In the
presence of the $G$-SSR this resource reduces to the {\em extractable} work
\beq
   W_G(\rho) \equiv W({\cal G}_G[\rho])\ .
   \label{W_G}
\eeq
The proof follows the same lines as that of Ref.~\cite{BarWis} for accessible entanglement. The crucial point here is
that once the work $W_G(\rho)$  has been extracted by a $G$-invariant operation, applying ${\cal G}$ to the system
does not change the amount of work that was extracted. According to \eqr{G invariant op}, the same result is obtained if
${\cal G}$ is applied to the system {\em before} the work is extracted, and so the extractable work is $W({\cal
G}_G[\rho])$. A {\em symmetric state}, i.e. one for which
\beq
   {\cal G}_G[\rho] = \rho\ ,
\eeq
suffers no loss in its ability to do work. In contrast, the extractable work possible for {\em
asymmetric} states (${\cal G}_G[\rho] \neq \rho$), is reduced under the $G$-SSR.

As an example, consider a spin-$\half $ particle prepared in state $\rho$ by Alice and sent to Bob, and let Bob have
knowledge only of the direction of Alice's $z$ axis. Bob cannot distinguish rotations by Alice about the $z$ axis. Thus
his knowledge of the state is constrained by the SSR induced by the $U(1)$ symmetry group associated with the unitary
representation
\beq
   \left\{T(\theta):\theta \in [0,2\pi),  T(\theta) = \exp(i \theta J_z/\hbar)  \right\}
\eeq
where $J_z = \frac{\hbar}{2}\sigma_z$ and  $\sigma_{z}$ is the Pauli operator for the $z$ component of spin.
Accordingly Bob ascribes the state
\beq
   {\cal G}_U[\rho] = \frac{1}{2\pi}\int_0^{2\pi}e^{i\theta J_{z}/\hbar}\rho e^{-i\theta J_{z}/\hbar}d\theta
\eeq
to the spin. Consider the spin-up state $\rho=\ket{1}\bra{1}$, where $\sigma_{z}\ket{\pm 1}=\pm
\ket{\pm 1}$.

The state $\ket{1}$ is symmetric with respect to $\{T(\theta)\}$ so for this state $W=1$ and the amount of extractable
work is also $W_{U}=1$. In contrast, the state $\ket{+}=(\ket{1}+\ket{-1})/\sqrt{2}$ is asymmetric with respect to
$\{T(\theta)\}$, with
\beq
   {\cal G}_U[\rho] = \half(\ket{1}\bra{1}+\ket{-1}\bra{-1})\ .
\eeq
Even though the state $\ket{+}$ has $W=1$, under the SSR Bob can  extract no work as $W_U=0$.

\subsection{Asymmetry}
A SSR thus introduces the need for a new resource: a system acting as a reference system to break the underlying
symmetry.  We now show that
\beq
   A_G (\rho)\equiv S({\cal G}_G[\rho]) - S(\rho)\ ,
   \label{A_G}
\eeq
which is the natural entropic measure of the {\em asymmetry} of $\rho$ with respect to $G$, is a measure that {\it
quantifies} the ability of a system to act as a reference system. To do this we need to show that $A_G$ has the following
properties:
\begin{list}{({\em\roman{enumi}})}{\usecounter{enumi}}
  \item  $A_G(\rho) \ge 0$;
  \item  $A_G(\rho) = 0$ iff $\rho$ is symmetric;
  \item  $A_G(\rho)$ cannot increase under the restriction of the $G$-SSR; and
  \item  $A_G(\rho)$ quantifies the ability of $\rho$ to act as a reference system.
\end{list}
The first two follow directly from the properties of the entropy function \cite{SWW}. For the
third, we have
\begin{theo}
  \label{thm_A_no_inc}
  No $G$-invariant operation can increase (on average) the asymmetry
  $A_G(\rho)$ of a state $\rho$.
\end{theo}

\begin{proof}
The most general $G$-invariant operation is a measurement that transforms an initial state $\rho$
into one of $M$ states
\beq
   \rho_j = \frac{1}{P_j}{\cal O}_j[\rho]\ ,
\eeq
such that
\beq
   {\cal O}_j[T(g) \rho T^\dagger(g)] = T(g) ({\cal O}_j[\rho]) T^\dagger(g), \forall g\in G\ ,
\eeq
with probability $P_j = \mbox{Tr}({\cal O}_j[\rho])$.This operation includes the possibility of adding ancillas in
prepared states and performing unitary operations and measurements on the combined system and ancillas. We wish to
show that
\beq
    A_G(\rho)\ge \sum_j P_j A_G(\rho_j)\ ,
\eeq
i.e. from \eqr{A_G}
\beq
  \label{ineq}
   S({\cal G}_G[\rho]) - S(\rho) \geq \sum_j P_j [ S({\cal G}_G[\rho_j]) - S(\rho_j) ]\ ,
\eeq
which can be rearranged as
\beq
   S({\cal G}_G[\rho]) - \sum_j P_j  S({\cal G}_G[\rho_j])\ \geq\ S(\rho) - \sum_j P_j S(\rho_j) \ .
   \label{ineq_rearranged}
\eeq
Note that, because the ${\cal O}_j$ are $G$-invariant, we can interchange the order of the twirl  ${\cal G}_G$ and the
${\cal O}_j$ operations.
Denoting the average change in entropy under the measurement operation by
\beq
  \langle \Delta S_{\cal O}(\rho) \rangle=S(\rho)- \sum_j P_j S(\rho_j)
\eeq
allows us to rewrite the inequality we wish to prove as
\beq
  \langle \Delta S_{\cal O}({\cal G}_G[\rho]) \rangle \geq \langle \Delta S_{\cal O}(\rho) \rangle\ .
  \label{ineq_final}
\eeq

We now use the following three facts:
\begin{list}{({\em\roman{enumi}})}{\usecounter{enumi}}
  \item for all operations the average entropy reduction, $\langle \Delta S_{\cal O}(\rho) \rangle$,
  is {\em concave} in $\rho$~\cite{KJ}, and so, e.g.
  $\ip{\Delta S_{\cal O}(\sum_j p_j\rho_j)}\ge\sum_jp_j\ip{\Delta S_{\cal O}(\rho_j)}$;

  \item the twirl operation produces the convex mixture
  \beq
     {\cal G}_G[\rho] = \frac{1}{|G|}\sum_g \sigma_g
  \eeq
  where $\sigma_g = T(g)\rho T^{\dagger}(g)$; and

  \item $\langle \Delta S_{\cal O}(\sigma_g)
  \rangle = \langle \Delta S_{\cal O}(\rho) \rangle$ for all $g$ due to the $G$-invariance of the ${\cal
  O}_j$ and the unitarity of the $T(g)$.
\end{list}
Putting these together we have
\beqa
  \langle \Delta S_{\cal O}({\cal G}_G[\rho]) \rangle
  &=& \langle \Delta S_{\cal O}\Big(\frac{1}{|G|}\sum_g \sigma_g\Big) \rangle\nn\\
  &\geq& \frac{1}{|G|}\sum_g \langle \Delta S_{\cal O}(\sigma_g) \rangle\nn\\
  &=& \frac{1}{|G|}\sum_g \langle \Delta S_{\cal O}(\rho) \rangle
  = \langle \Delta S_{\cal O}(\rho) \rangle\nn\\
\eeqa
which completes the proof of \eqr{ineq_final}.
\end{proof}

To show the fourth property let us first define $\Syn (X;\rho_1,\rho_2)$, the {\em synergy} of a quantity $X$, as
\beq
    \Syn (X;\rho_1,\rho_2)
    \equiv X(\rho_{1}\otimes\rho_{2}) - [X(\rho_{1}) +
    X(\rho_{2})]
    \label{DeltaX}
\eeq
for two systems in states $\rho_1$ and $\rho_2$. The extent to which system 1 acts as a reference system for system 2
(or vice versa) is the synergy of the extractable work, $\Syn(W_G; \rho_1,\rho_2)$; that is, the amount by which the
extractable work of the whole is larger than the sum of the extractable work of the parts. Then we have the following:

\begin{theo}
 \label{thm_A_ref}
The synergy of the extractable work is bounded by asymmetry:
\beq
   \Syn(W_G; \rho_1,\rho_2) \leq \mbox{\em min}\{ A_G(\rho_1), A_G(\rho_2)\}
\eeq
where $\rho_1$ and $\rho_2$ are arbitrary states of two systems sharing the same symmetry group $G$. Further, this
bound is achievable, in the sense that  for every $\rho_1$ there exists a $\rho_2$ such that $\Syn(W_G; \rho_1,\rho_2)
= A_G(\rho_1)$.
\end{theo}

\begin{proof}
We first note from Eqs.~(\ref{W}), (\ref{W_G}) and (\ref{A_G}) that the extractable work can be written as
\beq
  W_G(\rho)=W(\rho)-A_G(\rho)
  \label{WG_is_W_minus_A}
\eeq and, because $W(\rho_1\otimes\rho_2)=W(\rho_1)+W(\rho_2)$, the synergy of the extractable work
may be written as

\beqa
   &&\Syn (W_G; \rho_1,\rho_2)\nn\\
     &&\ = W_G(\rho_1\otimes\rho_2)-[W_G(\rho_1)+W_G(\rho_2)]\nn\\
     &&\ = [W(\rho_1\otimes\rho_2)-A_G(\rho_1\otimes\rho_2)]\nn\\
     &&\hspace{1cm}-[W(\rho_1)-A_G(\rho_1)+W(\rho_2)-A_G(\rho_2)]\nn\\
     &&\ = A_{G}(\rho_1) + A_{G}(\rho_2) - A_{G}(\rho_1\otimes\rho_2)\ .
   \label{DeltaW_G}
\eeqa
We next note that $A_{G}(\rho_1\otimes\rho_2)$ is equal to the Holevo $\chi$ quantity \cite{SWW},
$\chi_{12}$, for the ensemble
\beq
   \left\{ \left(P_g, \sigma_g\right)\ \forall g\in G\right\}
\eeq
where $P_g=|G|^{-1}$ is the probability associated with the state $\sigma_g$ and
\beq
      \sigma_g=[T(g) \rho_1 T(g)^\dagger] \otimes[T(g)\rho_2 T(g)^\dagger]\ .
\eeq
Similarly, the Holevo $\chi$ for the ensemble traced over subsystem 2 or 1 is $\chi_1 =
A_{G}(\rho_1)$ or $\chi_2 = A_{G}(\rho_2)$, respectively. The Holevo $\chi$ is non-increasing under
partial trace~\cite{SWW}, so
\beq
  A_{G}(\rho_1\otimes\rho_2)\geq A_{G}(\rho_j)\ \ {\rm for\ } j=1,2\ .
\eeq
Applying this to \eqr{DeltaW_G} gives the desired result.

To show achievability, choose $\rho_2=\ket{\psi}\bra{\psi}$ such that $\ip{\psi_{g'}|\psi_g} = \delta_{g,g'}$ where
$\ket{\psi_g}\equiv T_2(g)\ket{\psi}$ .  For finite groups this can be done with a normalisable state $\rho_2$ whereas
for Lie groups one can choose a normalisable state on a subspace of sufficiently large dimension~\cite{KitMayPre}.
Then using Eqs.~(\ref{inv_op}) and (\ref{T_factorisation}) we have
\beqa
   &&{\cal G}_G[\rho_1\otimes\rho_2])\nn\\
   &&\ \ = \frac{1}{|G|}\sum_{g\in G}
             [T_1(g)\otimes T_2(g)]\rho_1\otimes\rho_2[T^\dagger_1(g)\otimes T^\dagger_2(g)]\nn\\
   &&\ \ =  \frac{1}{|G|}\sum_{g\in G}
             [T_1(g)\rho_1T^\dagger_1(g)] \otimes \ket{\psi_g}\bra{\psi_g}\ .
\eeqa
The orthonormality of the set $\{\ket{\psi_g}:g\in G\}$ ensures that
\beqa
   S({\cal G}_G[\rho_1\otimes\rho_2])
   &=&\sum_{g\in G} \frac{1}{|G|}\left\{S[T_1(g)\rho_1 T^\dagger_1(g)]-\log(\frac{1}{|G|})\right\}\nn\\
   &=& S(\rho_1) + S({\cal G}_G[\rho_2])
\eeqa
where we have used $S_G(\rho_2)=\log(|G|)$. Finally, using this result with Eqs.~(\ref{A_G}) and
(\ref{DeltaW_G}) and noting that $S(\rho_1\otimes\rho_2)=S(\rho_1)+S(\rho_2)$ shows
\beqa
  && \Syn (W_G; \rho_1,\rho_2)  \nn\\
   &&\quad = \{S({\cal G}_G[\rho_1])-S(\rho_1)\} +\{S({\cal G}_G[\rho_2])-S(\rho_2)\}\nn\\
   &&\hspace{1cm}  -\left\{S({\cal G}_G[\rho_1\otimes\rho_2])-S(\rho_1\otimes\rho_2)\right\}\nn\\
     &&\quad= S({\cal G}_G[\rho_1]) + S({\cal G}_G[\rho_2]) - S({\cal G}_G\rho_1\otimes\rho_2])\nn\\
     &&\quad= S({\cal G}_G[\rho_1]) - S(\rho_1)\nn\\
     &&\quad= A_{G}(\rho_1)
\eeqa
which completes the proof of achievability.
\end{proof}

To illustrate the phenomenon of synergy, consider the previous spin-$\half$ example but now with {\em two} spins in
the state $\ket{+}$. That is, Alice sends to Bob the state $\rho_1\otimes \rho_2$, with $\rho_i = \ket{+}\bra{+}$ for
$i=1,2$. Bob again assigns the state
\beq
   {\cal G}_U[\rho_1\otimes \rho_2] =\frac{1}{2\pi}\int_0^{2\pi}e^{i\theta J_{z}/\hbar}(\rho_1\otimes \rho_2) e^{-i\theta J_{z}/\hbar}d\theta\ ,
\eeq
but now
\beq
    J_z = J_z^{\rm (1)} \otimes I^{\rm (2)} + I^{\rm (1)} \otimes J_z^{\rm (2)}
\eeq
where $I^{i}$ is the identity operator for system $i$.  We find
\beq
    {\cal G}_U[\rho_1\otimes \rho_2] =\uplus\smallfrac{1}{2} \ket{1,1} \uplus \smallfrac{1}{2} (\ket{1,-1} + \ket{-1,1}) \uplus
\smallfrac{1}{2} \ket{-1,-1}\ .
\eeq
Here, for clarity, we have used the following notational convention which was introduced in Ref.
\cite{JonWisBar}: $\uplus
\ket{\psi}$ is to be read as $+ \ket{\psi}\bra{\psi}$. Thus, for example,
\beq
   \uplus \alpha\ket{\psi} \uplus \beta\ket{\phi}
   \equiv |\alpha|^2 \ket{\psi}\bra{\psi}
   + |\beta|^2 \ket{\phi}\bra{\phi}\ .
\eeq
As before, $W(\rho_i)=1$, $W_{U}(\rho_i)=0$, and $A_{U}(\rho_i)=1$. But for the two spins together,
$W(\rho_1\otimes \rho_2)=2$, $W_{U}(\rho_1\otimes \rho_2)=\half$, and $A_{U}(\rho_1\otimes
\rho_2)=\smallfrac{3}{2}$. Thus the synergy is
\beq
    \Syn (W_{U}; \rho_1,\rho_2) = \half > 0\ .
\eeq
One spin acts as a reference for the other and partially breaks this $U(1)$-SSR.  Notice that the work synergy is less than
the asymmetries of the individual systems,  $\Syn (W_{U}; \rho_1,\rho_2) < A_{U}(\rho_i)=1$, in accord with
Theorem \ref{thm_A_ref}.

Having established the significance of $A_G(\rho)$ for indicating the ability of a system to act as a $G$-reference
system, we now observe that \eqr{WG_is_W_minus_A} represents a tradeoff or duality between this ability and the
amount of work that can be extracted under the $G$-SSR:
\beq
   W(\rho) = W_G(\rho) + A_G(\rho)\ .
   \label{comp_W}
\eeq
That is, under the $G$-SSR, the extractable work $W(\rho)$ represented by a given state is split into two new resources,
the extractable work $W_{G}$ and the asymmetry $A_{G}$.

\section{Extension to bipartite systems}

\subsection{Global and local SSRs}

Consider a system shared by two parties, Alice and Bob, such that the unitary representation of $G$
factorizes according to:
\beq
    T(g)=T_{\rm A}(g)\otimes T_{\rm B}(g)\ \forall\ g\in G\ .
    \label{T_is_TA_times_TB}
\eeq

There are two ways the $G$-SSR operates on the bipartite system, {\em globally} and {\em locally}. They can be
illustrated by considering their effect on the system state $\rho$. The global $G$-SSR acts when we have access to the
whole system using either non-local operations or transporting the whole system to one site. Thus in direct accord with
\eqr{inv_op} for the uni-partite case, our effective knowledge of the system under the global $G$-SSR is not $\rho$ but
\beqa
   {\cal G}_G[\rho] &=& \frac{1}{|G|}\sum_{g\in G}T(g)\rho T^\dagger(g)\nn\\
                    &=& \frac{1}{|G|}\sum_{g\in G}T_{\rm A}(g)\otimes T_{\rm B}(g)
                               \,\rho\, T^\dagger_{\rm A}(g)\otimes T^\dagger_{\rm B}(g)\ .\nn\\
\eeqa
In contrast, each party A and B has access only to the part of the system at their respective site. Accordingly the $G$-SSR
restricts their knowledge of the system to
\beq
   {\cal G}_{G\otimes G}[\rho] = \frac{1}{|G|^2}\sum_{g\in G}\sum_{g'\in G}
                     T_{\rm A}(g)\otimes T_{\rm B}(g')
                               \,\rho\, T^\dagger_{\rm A}(g)\otimes T^\dagger_{\rm B}(g')\ .
                               \label{twirledrho}
\eeq
We use the tensor product operator in the symbol ${\cal G}_{G\otimes G}$ to indicate that the twirl operation acts
locally on systems A and B; this is manifest in the sums over the independent indices $g$ and $g'$ in \eqr{twirledrho}.
We refer to the effect of ${\cal G}_{G\otimes G}$ as the {\em local} $G$-SSR.

The local $G$-SSR restricts the kinds of operations that the two parties can perform to local $G$-invariant operations
${\cal O}_{\rm AB}$ where
\beqa
    &&{\cal O}_{\rm AB}
       \left\{\left[T^{\phantom{\dagger}}_{\rm A}(g)\otimes T_{\rm B}(g')\right]
       \rho\left[T^\dagger_{\rm A}(g)\otimes T^\dagger_{\rm B}(g')\right]\right\}\nn    \\
    &&\ \ = T_{\rm A}(g)\otimes T_{\rm B}(g')\left\{{\cal O}_{\rm AB}
              [\rho]\right\} T^\dagger_{\rm A}(g)\otimes T^\dagger_{\rm B}(g')\ .\ \ \
              \label{local_G_O_AB}
\eeqa
for all $g, g'\in G$. This class includes (but is not limited to) products of local operations ${\cal O}_{\rm
A}\otimes{\cal O}_{\rm B}$, which could represent measurement outcomes. A wider class of allowed operations will
be defined below.  Moreover, any operation ${\cal O}_{\rm AB}$ which is local $G$-invariant is also global
$G$-invariant, because \eqr{local_G_O_AB} implies
\beq
  {\cal O}_{\rm AB}\left[T(g)\rho T^\dagger(g)\right]
  = T(g)\left\{{\cal O}_{\rm AB}[\rho]\right\} T^\dagger(g)
  \label{global_G_O_AB}
\eeq
for $T(g)=T_{\rm A}(g)\otimes T_{\rm B}(g)$.\\

\subsection{Globally-symmetric pure state $\rho^\beta$}

In this paper,  we restrict our analysis to globally symmetric pure states:
\beq
   \ket{\Psi}\bra{\Psi}={\cal G}_G\left[\ket{\Psi}\bra{\Psi}\right]\ .
\eeq
This requires $\ket{\Psi}$ to belong to a one-dimensional irrep of $G$. That is, using $\beta$ to
label the irrep,
\beq
   T(g)\ket{\Psi}=\lambda^\beta(g)\ket{\Psi} \ \ \forall\ g\in G
\eeq
where $T(g)$ is given by \eqr{T_is_TA_times_TB} and $\lambda^\beta(g)$ is the unit-modulus eigenvalue.

Let $G$ have $N_G$ distinct irreps $T^\mu(g)$ for $\mu=1,2, \cdots N_G$ and let Alice's operator
$T_{\rm A}(g)$ in \eqr{T_is_TA_times_TB} decompose into $K_{\rm A}$ irreps as
\beq
   T_{\rm A}(g) = \bigoplus_{n=1}^{K_{\rm A}}T^{f_{\rm A}(n)}(g) \ \ \forall\ g\in G
   \label{T_reduced}
\eeq
where
\beq
   f_{\rm A}(n) \in \{1,2, \cdots N_G\}
\eeq
labels an irrep for each $n$. The total number of irreps in $T_{\rm A}(g)$ can be written as $K_{\rm A} =
\sum_{\mu=1}^{N_G} M_{\rm A}^ \mu$, where $M_{\rm A}^\mu$ is the multiplicity (i.e. the number of copies) of
irreps of type $T^\mu$. The irrep $T^\mu$ operates on the $D_\mu$-dimensional subspaces spanned by
\cite{KitMayPre}
\beq
    \{\ket{\mu,m_\mu,i}: i=1,2,\cdots,D_\mu\}\ .
\eeq
The ``charge'' $\mu=1,2,\ldots N_G$ indexes the irreps $T^\mu$, the ``flavor'' $m_\mu=1,2,\ldots
M_{\rm A}^\mu$ indexes the copy of the irrep $T^\mu$ in the above decomposition, the ``color''
$i=1,2,\ldots D_\mu$ indexes an orthogonal basis set in which $T^\mu$ operates, and
\beq
    \ip{\mu,m_\mu,i|\nu,m_\nu,j} =\delta_{\mu,\nu}\delta_{m_\mu,m_\nu}\delta_{i,j}\ .
\eeq
Let Bob's operator $T_{\rm B}(g)$ have a similar decomposition. To find the form of the global $G$-invariant states
we need to consider pairs of conjugate irreps, that is pairs of irreps, say $T^\mu$ and $T^\nu$, whose tensor product
$T^\mu\otimes T^\nu$, can be reduced to a direct sum involving a given 1-dimensional irrep $T^\beta$ of $G$, i.e.
$T^\mu\otimes T^\nu\cong T^\beta\oplus\ldots$.  To do this we define $R^\beta$ to be the set of conjugate couples
\beq
   R^\beta=\{(\mu,\bar{\mu}):T^\mu(g)=C^\mu T^\beta(g)
     [T^{\bar{\mu}}(g)]^* (C^\mu)^{\dagger}\ \forall g\in G\}\label{R_beta}
\eeq
where $T^\beta$ is the given one-dimensional irrep  and $C^\mu$ is a unitary operator.  The entangled state
\beq
   \label{psi}
    \ket{\psi^{\mu,\beta}_{m_\mu,m_{\bar{\mu}}}}
      =\frac{1}{\sqrt{D_\mu}}\sum_{i,j}{C^\mu_{i,j}}\ket{\mu,m_\mu,i}\otimes\ket{\bar{\mu},m_{\bar{\mu}},j}
\eeq
for $(\mu,\bar{\mu})\in R^\beta$, is an eigenstate of $T(g)$ with eigenvalue
$\lambda^\beta(g)=T^\beta(g)$, and so it is globally symmetric.  The proof of this result is given
in Appendix \ref{app_global_symm}.

The most general, pure, globally symmetric state for a given value of $\beta$ is given by
\beq
   \rho^\beta=\uplus \ket{\Psi^\beta}
\eeq
where
\beq
  \ket{\Psi^\beta} =\sum_{\mu}\sum_{m_\mu,m_{\bar{\mu}}}
   d^{\mu}_{m_\mu,m_{\bar{\mu}}}
   \ket{\psi^{\mu,\beta}_{m_\mu,m_{\bar{\mu}}}}
\eeq
for arbitrary coefficients $d^{\mu}_{m_\mu,m_{\bar{\mu}}}$ satisfying
\beq
   \sum_{m_\mu,m_{\bar{\mu}}}
   |d^{\mu}_{m_\mu,m_{\bar{\mu}}}|^2=1\ .
\eeq
In the following we evaluate the effect of the $G$-SSR on the resources represented by this general state $\rho^\beta$.

\subsection{Unconstrained entanglement of $\rho^\beta$}

We begin by evaluating the total entanglement in $\rho^\beta$, measured in terms of the entropy of entanglement,
without the restriction of the $G$-SSR.  It is convenient to factorize the representation into ``flavor'' (indexed by
$m_\mu$) and ``color'' (indexed by $i$) subsystems as
\beq
    \ket{\mu,m_\mu,i}\equiv\ket{\mu,m_\mu}\ket{\mu,i}
    \label{charge_flavour}
\eeq
and rewrite the state in terms of states of the flavor and color subsystems as
\beq
  \ket{\Psi^\beta} =\sum_{\mu}\sqrt{P_\mu}\ket{\varphi_\mu}
\Big(\frac{1}{\sqrt{D_\mu}}\sum_{i,j}{C^\mu_{i,j}}\ket{\mu,i}\otimes\ket{\bar{\mu},j}\Big)
\eeq
where
\beqa
  \ket{\varphi_\mu}
    &=& \sum_{m_\mu,m_{\mu}}
     \frac{d^\mu_{m_\mu,m_{\bar{\mu}}}}{\sqrt{P_\mu}}\ket{\mu,m_\mu}\otimes\ket{\bar\mu,m_{\bar\mu}}\\
     \label{flavor_subsystem}
    P_\mu&=& \sum_{m_\mu,m_{\bar{\mu}}} |d^\mu_{m_\mu,m_{\bar{\mu}}}|^2\  .
\eeqa
Then taking the partial trace of $\rho^\beta$ over Bob's state space and making use of the unitarity of $C^\mu$ yields
\beq
   {\rm Tr}_{\rm B}(\rho^\beta)
   =\sum_\mu  P_\mu\ro{ \biguplus_{k}
   \Lambda^\mu_k\ket{{\rm A}^\mu_k}} \otimes \ro{
   \biguplus_i  \frac{1}{\sqrt{D_\mu}}\ket{\mu,i}}\ ,
\eeq
where
\beq
   \biguplus_i \ket{\psi_i} \equiv \sum_i \uplus \ket{\psi_i} =  \sum_i \ket{\psi_i}\bra{\psi_i}\ ,
\eeq
and we have used the Schmidt decomposition
\beq
   \ket{\varphi_\mu}
       =\sum_{k}\Lambda^\mu_k\ket{{\rm A}^\mu_k}\otimes\ket{{\rm
         B}^\mu_k}\label{SchmidtDecomposition}
\eeq
of the state of the bipartite flavor subsystem with
\beq
  \ip{{\rm A}^\mu_k|{\rm A}^\mu_l}=\ip{{\rm B}^\mu_k|{\rm B}^\mu_l}=\delta_{k,l}\ .
\eeq
Thus the entanglement is given by
\beq
   E(\rho^\beta)=-\sum_{\mu,k}P_\mu|\Lambda^\mu_k|^2
   \log(P_\mu|\Lambda^\mu_k|^2)+\sum_\mu P_\mu \log (D_\mu)\ .
\eeq

\subsection{Resources in $\rho^\beta$ under the local $G$-SSR}

\subsubsection{Entanglement accessible under local $G$-SSR} The accessible entanglement in the state $\ket{\Psi^\beta}$
constrained by the local $G$-SSR is, according to \cite{BarWis}, given by the total entanglement in the state
\cite{GlobalNote}.
\beq
\label{twirledrhobeta}
  {\cal G}_{G\otimes G} [\rho^\beta]
  = \frac{1}{|G|^2}\biguplus_{g,g'\in G}  \left([T_{\rm A}(g)
       \otimes T_{\rm B}(g')] \ket{\Psi^\beta}\right)\ .
\eeq
where ${\cal G}_{G\otimes G}$ is defined in \eqr{twirledrho}.  Using the unitarity of the matrices $C^\mu$ and the
grand orthogonality theorem \cite{ortho}
\beq
  \sum_{g\in G} T^\mu_{k,l}(g)[T^{\nu}_{n,m}(g)]^\ast =\frac{|G|}{D_\mu}\delta_{\mu,\nu}\delta_{k,n}\delta_{l,m}
  \label{ortho_theorem}
\eeq
where $T^{\eta}_{i,j}=\ip{\eta,m_\eta,i|T^\eta|\eta,m_\eta,j}$ \cite{ortho}, yields
\beqa
  &&{\cal G}_{G\otimes G} [\rho^\beta] =\sum_\mu P_\mu\big(\uplus \ket{\varphi_\mu}\big)\nn\\
  &&\hspace{2.5cm}   \otimes\biguplus_{i,j}
   \ro{ \frac{1}{D_\mu}{\ket{\mu,i}\otimes\ket{\bar{\mu},j}}}\ .\label{local_G}
\eeqa
The entropy of this state is easily found using \eqr{twirledrhobeta} and \eqr{ortho_theorem} to be
\beq
     S({\cal G}_{G\otimes G} [\rho^\beta]) = H_{G\otimes G}^{\rm (ch)}(\rho^\beta)
     + 2H_{G\otimes G}^{\rm (co)}(\rho^\beta)
     \label{S_GxG=Hch+Hco}
\eeq
where we have defined color and charge correlations
\beqa
  H_{G\otimes G}^{\rm (co)}(\rho^\beta)&\equiv&\sum_\mu P_\mu \log D_\mu\ ,
  \label{H_Gco}\\
  H_{G\otimes G}^{\rm (ch)}(\rho^\beta)&\equiv&-\sum_\mu P_\mu \log P_\mu\ .
  \label{H_Gch}
\eeqa

We note that Alice (or, equivalently, Bob) can make a measurement of the charge without changing the amount of
accessible entanglement, because the measurement commutes with all $G$-invariant operations \cite{WisVac,BarWis}.
The proof can be found in Appendix \ref{app_charge_meas}.

This local measurement of charge yields the value of $\mu$ with probability $P_\mu$, resulting in the pure entangled
state $\ket{\varphi_\mu}$ of the flavor subsystem. The entanglement in the flavor subsystem is then the entropy
$-\sum_k |\Lambda^\mu_k|^2\log (|\Lambda^\mu_k|^2)$ of Alice's reduced state, $\biguplus_k (\Lambda^\mu_k
\ket{{\rm A}^\mu_k})$.  The corresponding state of the color subsystem in \erf{local_G} is
$\biguplus_{i,j}\ro{\ket{\mu,i}\otimes\ket{\bar{\mu},j}/D_\mu }$ which is clearly separable, and so  the color
subsystem makes no contribution to the entanglement. Averaging over all  $\mu$ values gives the {\em accessible
entanglement $E_{G\otimes G}$ under local $G$-SSR} as
\beq
  E_{G\otimes G}(\rho^\beta)=E(\rho^\beta)-H_{G\otimes G}^{\rm (co)}(\rho^\beta)
       -H_{G\otimes G}^{\rm (ch)}(\rho^\beta)\ .
    \label{E_G}
\eeq
The quantity  $E_{G\otimes G}(\rho^\beta)$ in (\ref{E_G}) represents the ability of the system under the local
$G$-SSR to do ``logical work'' in the form of bipartite quantum information processing \cite{OppHor}.

\subsubsection{Work extractable under local $G$-SSR and LOCC}

Just as in the unipartite case in the absence of the $G$-SSR, a bipartite state $\rho$ can be used to extract mechanical
work {\em locally} at each site from local thermal reservoirs \cite{OppHor,HorOpp03}. Only local operations and
classical communication (LOCC) are allowed for the extraction process which results in a maximum amount of work
$W_{\rm L}(\rho)$ being extracted in total. Oppenheim {\em et al.} \cite{OppHor} showed that the quantum
correlations in the state $\rho$ reduce the amount of work that can be extracted in this way. Alternatively one could
transmit the system at Alice's site to Bob's site through a dephasing channel and then extract the work locally at Bob's
site. They showed that for pure states $\rho$ an equivalent amount of work $W_{\rm L}(\rho)$ is obtained if the
dephasing channel produces a classically correlated state of minimum entropy. The allowed operations $\cal Q$ for this
method are those that can be realized using local unitaries, local ancillas (whose extractable work must be subtracted off
at the end) and transmission through the dephasing channel. That is \cite{OppHor}
\beq
  W_{\rm L}(\rho)=W(\widetilde{\cal Q}[\rho])\label{W_Q_P}
\eeq
where $ \widetilde{\cal Q}$ is the optimum allowed operation that yields a classically-correlated
state with minimal entropy $S(\widetilde{\cal Q}[\rho])$. For pure states there is a duality
between abilities to do mechanical and ``logical'' work \cite{OppHor}:
\beq
   W(\rho) = W_{\rm L}(\rho) + E(\rho)\ .\label{W_is_Wlo+E}
\eeq
In particular, consider the locally extractable mechanical work from the pure state
\beq
   \sigma=\uplus \big(\sum_{n}\sqrt{p_n}\ket{\phi_n}_{\rm AB} \otimes\ket{\chi_n}_{\rm AB}
    \otimes\ket{\psi_n}_{\rm AB} \big)
\eeq
in the absence of the $G$-SSR. Here the $\ket{\phi_m}$, $\ket{\chi_m}$ and $\ket{\psi_m}$ represent states of three
bipartite systems satisfying
\beq
  \ip{\phi_n|\phi_m}=\ip{\chi_n|\chi_m}=\ip{\psi_n|\psi_m}=\delta_{n,m}\ .
\eeq
From \cite{OppHor} the optimum operation $\widetilde{\cal Q}$ dephases $\sigma$ in its Schmidt basis; this can be
carried out by first dephasing in the Schmidt basis of $\{\ket{\chi_n}\otimes\ket{\psi_n}\}$ followed by dephasing in
the Schmidt basis of $\{\ket{\phi_n}\}$. Let the Schmidt bases be given by
\beqa
    \ket{x_n} &=& \sum_i x_{n,i}\ket{x_{n,i}}
\eeqa
where $x_{n,i}$ are the Schmidt coefficients and $\ket{x_{n,i}}$ are a set of orthonormal states,  for $x$ being $\phi$,
$\psi$ or $\chi$. The first step yields a state of the form
\beqa
   \sigma^\prime &=&\sum_{n}p_n
    \big(\uplus\ket{\phi_n}_{\rm AB}\big)\nn\\
    &&\otimes\biguplus_{i,j} \big(
   \chi_{n,i}\ket{\chi_{n,i}}_{\rm AB} \otimes\psi_{n,j}\ket{\psi_{n,j}}_{\rm AB} \big)
   \label{sigma_prime}
\eeqa
and the second step yields a state of the form
\beqa
   \sigma^{\prime\prime}&=&\sum_{n}p_n \biguplus_{k}\big(\phi_{n,k}\ket{\phi_{n,k}}_{\rm AB}\big)\nn\\
         &&\otimes\biguplus_{i,j} \big(\chi_{n,i}\ket{\chi_{n,i}}_{\rm AB}
         \otimes\psi_{n,j}\ket{\psi_{n,j}}_{\rm AB} \big)\ .
   \label{sigma_prime_prime}
\eeqa
This can be {\em reversibly} transformed using the dephasing channel into
\beqa
   &&\sum_{n}p_n \biguplus_{k}\big(\phi_{n,k}\ket{\phi_{n,k}}_{\rm BB}\big)\nn\\
   &&\quad \otimes\biguplus_{i,j} \big( \chi_{n,i}\ket{\chi_{n,i}}_{\rm BB}
   \otimes\psi_{n,j}\ket{\psi_{n,j}}_{\rm BB} \big)\ ,
\eeqa
where the whole system is located at site B. The maximum amount of mechanical work  that can be extracted from
$\sigma$ locally at each site is equal to the maximum that can be extracted locally at site B from
$\sigma^{\prime\prime}=\widetilde{\cal Q}[\sigma]$, as given in \eqr{W_Q_P}. We use this result below.

We now consider the local mechanical work $W_{G\otimes G-{\rm L}}(\rho)$ that is extractable from state $\rho$
under both the local $G$-SSR and the LOCC restrictions. This is given by

\beqa
  W_{G\otimes G-{\rm L}}(\rho)
  =W_{\rm L}({\cal G}_{G\otimes G}[\rho])
  = W(\widetilde{\cal Q}_{G\otimes G}
      \{{\cal G}_{G\otimes G}[\rho]\})\ , \nn\\
\eeqa
where $\widetilde{\cal Q}_{G\otimes G}$ is locally $G$-invariant.

We first evaluate $W(\widetilde{\cal Q}_{G\otimes G}\{{\cal G}_{G\otimes G} [\rho^\beta]\})$. As ${\cal
G}_{G\otimes G} [\rho^\beta]$ in \erf{local_G} is equivalent in form to $\sigma^\prime$ in \eqr{sigma_prime}, the
optimum operation $\widetilde{\cal Q}_{G\otimes G}$ is dephasing in the Schmidt basis of the flavor subsystem
$\{\ket{\varphi_\mu}\}$ in \eqr{flavor_subsystem}. This can be done by making a {\em local} measurement in the
Schmidt basis given in Eqs.~(\ref{SchmidtDecomposition}) and (\ref{flavor_subsystem}). This operation can be shown
to be local $G$-invariant as follows.  For example, a local measurement by Alice that projects onto the Schmidt basis is
described by the set of projection operators
\beq
    \widetilde{\Pi}_k =\Big(\ket{{\rm A}^\mu_k}\bra{{\rm A}^\mu_k}\otimes\openone^{\rm (co)}_\mu\Big)_{\rm A}
       \otimes \openone_{\rm B}
\eeq
for the same set of values of $k$ as in \eqr{SchmidtDecomposition} and where $\openone^{\rm (co)}_\mu =
\sum_{i=1}^{D_\mu}\ket{\mu,i}\bra{\mu,i}$ projects onto the color subsystem. Recalling the decomposition
\eqr{T_reduced} and noting that the irrep $T^\mu(g)$ acts on the corresponding color subsystem only, we find
\beqa
    &&\hspace{-10mm}\Big[T_{\rm A}(g)\otimes T_{\rm B}(g')\Big]\widetilde{\Pi}_k\nn\\
                    &=&   T_{\rm A}(g)\otimes T_{\rm B}(g')\Big[\Big(\ket{{\rm A}^\mu_k}\bra{{\rm A}^\mu_k}
                                \otimes \openone^{\rm (co)}_\mu\Big)_{\rm A} \otimes \openone_{\rm B}\Big]\nn\\
                    &=&   \Big[\ket{{\rm A}^\mu_k}\bra{{\rm A}^\mu_k}\otimes T^\mu(g)\Big]_{\rm A}
                           \otimes T_{\rm B}(g')\nn\\
                    &=&   \Big[\Big(\ket{{\rm A}^\mu_k}\bra{{\rm A}^\mu_k}\otimes \openone^{\rm (co)}_\mu\Big)_{\rm A}
                           \otimes \openone_{\rm B}\Big]\Big[T_{\rm A}(g)\otimes T_{\rm B}(g')\Big]\nn\\
                    &=& \widetilde{\Pi}_k \Big[T_{\rm A}(g)\otimes T_{\rm B}(g')\Big]\ .
\eeqa
That is, projection by $\widetilde{\Pi}_k$ is a locally $G$-invariant operation according to \eqr{local_G_O_AB}. The
average result of Alice's measurement gives the desired optimal dephasing, i.e.
\beqa
   &&\hspace{-5mm} \widetilde{\cal Q}_{G\otimes G}\{{\cal G}_{G\otimes G} [\rho^\beta]\}\nn\\
   &=& \sum_k p_k\widetilde{\Pi}_k\, \rho^\beta\,\widetilde{\Pi}_k\nn\\
   &=& \biguplus_{\mu,k,i,j} \left( \frac{1}{D_\mu}\sqrt{P_\mu}\Lambda^\mu_k \ket{{\rm
    A} ^\mu_k}\otimes\ket{{\rm B}^\mu_k}\otimes \ket{\mu,i}\otimes\ket{\bar{\mu},j} \right)\ ,\nn\\
\eeqa
where $p_k={\rm Tr}(\widetilde{\Pi}_k \rho^\beta)$.

Now using \eqr{W_Q_P} the locally extractable work is found to be
\beqa
   W_{G\otimes G-{\rm L}}(\rho^\beta)
   &=& W(\widetilde{\cal Q}_{G\otimes G}\{{\cal G}_{G\otimes G} [\rho^\beta]\})\nn\\
   &=&  \ln(D)-S(\widetilde{\cal Q}_{G\otimes G}\{{\cal G}_{G\otimes G} [\rho^\beta]\})\nn\\
   &=& \log D - [E(\rho^\beta) +  H^{\rm (co)}_{G\otimes G}(\rho^\beta)]\nn\\
   \label{W_GxG_is_lnD_E_Hco}
\eeqa
The global symmetry of the pure state $\rho^\beta$ ensures that
\beq
     W_G(\rho^\beta) = W({\cal G}_G[\rho^\beta]) = W(\rho^\beta)=\log D
\eeq
and so the locally extractable work can be written as
\beq \label{WGlo}
   W_{G\otimes G-{\rm L}}(\rho^\beta) = W(\rho^\beta) - [E(\rho^\beta)
   + H^{\rm (co)}_{G\otimes G}(\rho^\beta)]\ .
\eeq
Finally, using \eqr{W_is_Wlo+E}  we can rearrange \eqr{WGlo} as
\beq
   W_{G\otimes G-{\rm L}}(\rho^\beta) = W_{\rm L}(\rho^\beta) - H^{\rm (co)}_{G\otimes G}(\rho^\beta)
\eeq
which shows that the reduction in $W_{\rm L}$  due to the local $G$-SSR is manifest
in the mixing in the color subsystems.\\

\subsubsection{Shared asymmetry with respect to local $G$-SSR}

Eqs.~(\ref{E_G}) and (\ref{WGlo}) show that under the local $G$-SSR the duality between logical and local
mechanical work in \eqr{W_is_Wlo+E} is broken, i.e. $W \ne W_{G\otimes G-{\rm L}} + E_{G\otimes G}$. Just as
in the unipartite case, the lack of a reference system results in the loss of the ability to do work. In this
globally-symmetric bipartite case what is lacking is a {\em shared} reference system.  For a globally symmetric system to
act as a shared reference there must be correlations (quantum or classical) between the asymmetries for each party. These
correlations are unaffected by a global transformation ${\cal G}$, but are destroyed by local mixing ${\cal
G}_{G\otimes G}$. This suggests the natural entropic measure for the ability of such a system to act as a shared
reference is
\beq
  A_{G\otimes G}^{\rm (sh)}(\rho)
  \equiv S({\cal G}_{G\otimes G}\{{\cal G}_G[\rho]\})-S({\cal G}_G[\rho])\label{A_sh}
\eeq
which we call the {\em shared asymmetry}.  Notice that here $\rho$ is an arbitrary state which is not necessarily pure nor
globally symmetric.  By shared we mean that both Alice and Bob have access to this type of asymmetry for unlocking the
resources represented by $\rho$ at their sites. The global asymmetry $A_G[\rho]$ of the state is not, in itself, useful for
this purpose. To eliminate the effects of {\em global} asymmetry we have defined  $A_{G\otimes G}^{\rm (sh)}(\rho)$
in \eqr{A_sh} in terms of the globally-symmetric state ${\cal G}_G[\rho]$.  The result is that $A_{G\otimes G}^{\rm
(sh)}(\rho)$ is equal to the increase in entropy due to the local $G$-SSR only. For the U(1) case the {\em refbit}
\cite{Enk05} has $A_{G\otimes G}^{\rm (sh)}=1$, as one would like.

In analogy with $A_G$ for the unipartite case, we now show that $A_{G\otimes G}^{\rm (sh)}$ similarly quantifies the
resource of acting as a shared reference system for arbitrary states $\rho$.  First we note that

\begin{widetext}
\beqa
    {\cal G}_{G\otimes G}\{{\cal G}_G[\rho]\}
    &=& \frac{1}{|G|^2}\sum_{g,g'\in G}\frac{1}{|G|}\sum_{g''\in G}[T(g')\otimes T(g)][T(g'')\otimes T(g'')]\rho
                    [T\dg(g'')\otimes T\dg(g'')][T\dg(g')\otimes T\dg(g)]\nn\\
    &=& \frac{1}{|G|^2}\sum_{g,g'\in G}\frac{1}{|G|}\sum_{g''\in G}[T(g'\circ g'')\otimes T(g\circ g'')]\rho
                    [T\dg(g'\circ g'')\otimes T\dg(g\circ g'')]\nn\\
    &=& \frac{1}{|G|^2}\sum_{g,g'\in G}\frac{1}{|G|}\sum_{h\in G}[T(g'\circ g^{-1}\circ h)\otimes T(h)]\rho
                    [T\dg(g'\circ g^{-1}\circ h)\otimes T\dg(h)]\nn\\
    &=& \frac{1}{|G|}\sum_{h'\in G}\frac{1}{|G|}\sum_{h\in G}[T(h')T(h)\otimes T(h)]\rho
                    [T\dg(h)T\dg(h')\otimes T\dg(h)]\label{G_times_G_G}\nn\\
    &=& \frac{1}{|G|}\sum_{h'\in G}\frac{1}{|G|}\sum_{h\in G}[T(h')\otimes \openone][T(h)\otimes T(h)]\rho
                    [T\dg(h)\otimes T\dg(h)][T\dg(h')\otimes \openone]\nn\\
    &=& {\cal G}_{G\otimes \{e\}}\{{\cal G}_G[\rho]\}= {\cal G}_{\{e\}\otimes G}\{{\cal G}_G[\rho]\}
\eeqa
\end{widetext}
where $h=g\circ g''$, $h'=g'\circ g^{-1}$ and $\{e\}$ is the group containing only the identity element so that, for
example,
\beqa
    {\cal G}_{G\otimes \{e\}}[\rho]
        &=&\frac{1}{|G|}\sum_{g\in G}[T_{\rm A}(g)\otimes \openone_{\rm B}]\rho
            [T\dg_{\rm A}(g)\otimes \openone_{\rm B}]\label{G_times_e} \ .\nn\\
\eeqa
Similarly we can show
\beq
    {\cal G}_{G\otimes G}\{{\cal G}_G[\rho]\}
    = {\cal G}_{\{e\}\otimes G}\{{\cal G}_G[\rho]\}
\eeq
and this means that the shared asymmetry may be written equivalently as
\beqa
   A_{G\otimes G}^{\rm (sh)}(\rho)
     &=& S({\cal G}_{G\otimes G} [\rho])- S({\cal G}_G[\rho])\ .
\eeqa

The properties of the entropy function show  $A_{G\otimes G}^{\rm (sh)}$ has the following two properties:
\begin{list}{({\em\roman{enumi}})}{\usecounter{enumi}}
   \item $A_{G\otimes G}^{\rm (sh)}(\rho) \geq 0$~;
   \item $A_{G\otimes G}^{\rm (sh)}(\rho) = 0$ iff ${\cal G}_{G\otimes G}[\rho] = {\cal G}_{G}[\rho]$.
\end{list}
For global $G$-invariant states $\rho^\beta$ we have ${\cal G}_G[\rho^\beta]=\rho^\beta$ and so the second property
becomes
\begin{list}{({\em ii})$^\prime$}{}
   \item $A_{G\otimes G}^{\rm (sh)}(\rho^\beta) = 0$ iff ${\cal G}_{G\otimes G}[\rho^\beta] =
   \rho^\beta$, or, equivalently, iff ${\cal G}_{G\otimes \{e\}}[\rho^\beta]
    = {\cal G}_{\{e\}\otimes G}[\rho^\beta]=\rho^\beta$\ .
\end{list}

A third property is
\begin{list}{({\em\roman{enumi}})}{\usecounter{enumi}\setcounter{enumi}{2}}
    \item $A_{G\otimes G}^{\rm (sh)}(\rho)$ is non-increasing on average under {\em local} $G$-invariant local
    operations and classical communication (LOCC),
\end{list}
which is analogous to the third property of $A_G$.  We define local $G$-invariant LOCC as those LOCC that are
allowed by the local $G$-SSR or, equivalently, those that satisfy \eqr{local_G_O_AB}. These include products of local
operations of the form ${\cal O}_{\rm A}\otimes{\cal O}_{\rm B}$. For classical communication to be permitted
under the $G$-SSR, the information must be carried by physical processes that are permitted by the $G$-SSR. This can
be done, for example, by using a $G$-SSR--free system as the carrier, which we assume is the case. The class of LOCC
is a subset of the class of {\it separable} operations \cite{BennDiVi}. It is straight-forward to show, using the same
reasoning as in Ref.~ \cite{BennDiVi}, that every local $G$-invariant LOCC operation is a local $G$-invariant
separable operation. So to prove the third property it is sufficient to show that $A_{G\otimes G}^{\rm (sh)}(\rho)$ is
non-increasing under local $G$-invariant {\em separable} operations of the type
\beq
    \{{\cal O}_{{\rm A},i}\otimes{\cal O}_{{\rm B},j}:i=1,2,\ldots, j=1,2,\ldots,\}
    \label{local_G_invariant_sep}
\eeq
where  ${\cal O}_{{\rm A},i}\otimes{\cal O}_{{\rm B},j}$ satisfies \eqr{local_G_O_AB}. We therefore wish to show
that
\beq
    A_{G\otimes G}^{\rm (sh)}(\rho)\ge \sum_{i,j} P_{i,j} A_{G\otimes G}^{\rm (sh)}(\rho_{i,j})
\eeq
where
\beqan
    \rho_{i,j}&=&\frac{1}{P_{i,j}}\left({\cal O}_{{\rm A},i}\otimes{\cal O}_{{\rm B},j}\right)[\rho]\\
    P_{i,j}&=&{\rm Tr}\left[\left({\cal O}_{{\rm A},i}\otimes{\cal O}_{{\rm
    B},j}\right)\rho\right]\ ,
\eeqan
or equivalently, from \eqr{A_sh},
\beqa
   &&S({\cal G}_{G\otimes G}\{{\cal G}_{G}[\rho]\}) - S({\cal G}_{G}[\rho])\non\\
   &&\ \geq \sum_{i,j} P_{i,j}\Big[S({\cal G}_{G\otimes G}\{{\cal G}_{G}[\rho_{i,j}]\})
        - S({\cal G}_{G}[\rho_{i,j}]) \Big].\quad\quad
   \label{A_sh_3rd_prop}
\eeqa
We note that according to Eqs.~(\ref{local_G_O_AB}) and (\ref{global_G_O_AB}) each element in the set in
\eqr{local_G_invariant_sep} is also global $G$-invariant, and so
\beq
    {\cal G}_G\left\{({\cal O}_{{\rm A},i}\otimes{\cal O}_{{\rm B},j})[\rho]\right\}
       = ({\cal O}_{{\rm A},i}\otimes{\cal O}_{{\rm B},j})
              \{{\cal G}_G[\rho]\}\ .
\eeq
This means we can interchange the twirl and measurement operations in \eqr{A_sh_3rd_prop}.  Let $ \varrho \equiv
{\cal G}_{G}[\rho]$ and
\beqa
    \varrho_{i,j} \equiv \frac{1}{P_{i,j}}({\cal O}_{{\rm A},i}\otimes{\cal O}_{{\rm B},j})\left\{{\cal
        G}_{G}[\rho]\right\} = {\cal G}_{G}[\rho_{i,j}]\ .
\eeqa
We can now rewrite \eqr{A_sh_3rd_prop} as
\beqa
   S({\cal G}_{G\otimes G}[\varrho]) - S(\varrho)
   \ \geq \sum_{i,j} P_{i,j}\Big[S({\cal G}_{G\otimes G}[\varrho_{i,j}])
        - S(\varrho_{i,j}) \Big]\nn\\
   \label{A_sh_3rd_prop_equiv}
\eeqa
which is in the same form as \erf{ineq}. The same arguments which follow \erf{ineq} can be used to
show that the right hand side of \erf{A_sh_3rd_prop_equiv}, and thus $A_{G\otimes G}^{\rm
(sh)}(\rho)$, is non-increasing under local $G$-invariant {\em separable} operations, and by
implication, that the third property is therefore valid.\\

The {\em total} amount of extractable work under the local $G$-SSR is the sum of the logical work and the locally
extracted mechanical work, i.e. $(W_{G\otimes G-{\rm L}} + E_{G\otimes G})$.   Using
Eqs.~(\ref{S_GxG=Hch+Hco}), (\ref{E_G}) and (\ref{W_GxG_is_lnD_E_Hco})  we find
\beqa
   &&W_{G\otimes G-{\rm L}}(\rho^\beta) + E_{G\otimes G}(\rho^\beta)\nn\\
    &&\hspace{1cm}= \log D-H_{G\otimes G}^{\rm (ch)}(\rho^\beta)
                                - 2H_{G\otimes G}^{\rm (co)}(\rho^\beta) \nn\\
       &&\hspace{1cm}= \log D - S({\cal G}_{G\otimes G}[\rho^\beta])\ .
   \label{W_total_logical_mech}
\eeqa
This total is equivalent to just the mechanical work  $W_{G\otimes G}(\rho^\beta)$, where we define
\beq
   W_{G\otimes G}(\rho) \equiv \log D - S({\cal G}_{G\otimes G}[\rho])
    \label{W_GxG}
\eeq
for arbitrary states $\rho$ (which are not necessarily globally symmetric). The physical interpretation of $W_{G\otimes
G}(\rho)$ is that under the local $G$-SSR, ${\cal G}_{G\otimes G}(\rho)$ is the effective state of the system which
can be transferred locally to SSR-free ancillas at each site. Once the transfer is done the amount of work that can be
extracted {\em globally} from a thermal reservoir (i.e. without the LOCC restriction) using the ancillas is
$W_{G\otimes G}(\rho)$.  Eqs.~(\ref{W_total_logical_mech}) and (\ref{W_GxG}) show that an equivalent physical
interpretation of $W_{G\otimes G}(\rho)$ is that it is the total extractable work, both logical and mechanical, that can
be extracted under the local $G$-SSR and LOCC. This result leads to the fourth and final property that
\begin{list}{({\em\roman{enumi}})}{\usecounter{enumi}\setcounter{enumi}{3}}
   \item the shared asymmetry is an achievable upper bound on the
    synergy of the total extractable work $W_{G\otimes G}$.
\end{list}
This follows from the following theorem.

\begin{theo}
    \label{thm_A_sh_ref}
    The synergy of the
    total work $W_{G\otimes G}$ under the local $G$-SSR is bounded by the shared asymmetry, i.e.
    \beq
       \Syn(W_{G\otimes G}^{\rm (tot)}; \rho_1,\rho_2) \leq
          \mbox{\em min}\{A_{G\otimes G}^{\rm (sh)}(\rho_1),A_{G\otimes G}^{\rm (sh)}(\rho_2)\}\ .
    \eeq
    The upper bound is achievable in the sense of theorem~\ref{thm_A_ref}.
\end{theo}

We omit the proof, which has the same form as that of Theorem~\ref{thm_A_ref}. The achievability
follows from the existence of bipartite globally symmetric states $\ket{\Psi}$ such that
\beq
  \bra{\Psi}T_{\rm A}(g)\dg T_{\rm A}(g') \ket{\Psi} = \delta_{g,g'}\ .
\eeq

Thus we have identified three resources that emerge in a bipartite setting under a $G$-SSR: the
locally extractable mechanical work $W_{G\otimes G-{\rm L}}$, the accessible entanglement or
logical work $E_{G\otimes G}$, and the shared asymmetry $A_{G\otimes G}^{\rm (sh)}$. Finally, we
show that, for globally $G$-invariant states, there is a triality relation between them,
generalizing the duality (\ref{comp_W}) from the unipartite setting. A straightforward calculation
gives
\beq
     A_{G\otimes G}^{\rm (sh)}(\rho^\beta) = 2H_{G\otimes G}^{\rm (co)}(\rho^\beta)+H_{G\otimes G}^{\rm (ch)}(\rho^\beta)\ ,
\eeq
which, together with \erf{E_G} and \erf{WGlo}, then gives the main result of this paper
\begin{equation}
  W_{G}(\rho^\beta)=W_{G\otimes G-{\rm L}}(\rho^\beta)
  +E_{G\otimes G}(\rho^\beta) +A_{G\otimes G}^{\rm (sh)}(\rho^\beta)\ .
  \label{comp_E}
\end{equation}

\subsubsection{Local asymmetry with respect to local $G$-SSR}

We defined the shared asymmetry of state $\rho$ in \eqr{A_sh} as the extra entropy generated by the local $G$-SSR
acting on the state ${\cal G}_G[\rho]$, i.e. $A^{\rm (sh)}_{G\otimes G}(\rho)=S({\cal G}_{G\otimes G}\{{\cal
G}_G[\rho]\})- S({\cal G}_G[\rho])$. It is interesting to consider the entropy generated by the local $G$-SSR acting on
the state $\rho$ itself.  For this purpose we define
\beq
   A^{\rm (lo)}_{G\otimes G}(\rho)\equiv S({\cal G}_{G\otimes G}[\rho])-S(\rho)\ ,
   \label{A_lo}
\eeq
which we call the {\em local asymmetry} of $\rho$. $A^{\rm (lo)}_{G\otimes G}$ is related to the
shared $A^{\rm (sh)}_{G\otimes G}$ and global $A_{G}(\rho)$ asymmetries by
\beqa
      A^{\rm (sh)}_{G\otimes G}(\rho) &=& A^{\rm (lo)}_{G\otimes G}({\cal G}_G[\rho])\\
      A^{\rm (lo)}_{G\otimes G}(\rho) &=& A^{\rm (sh)}_{G\otimes G}(\rho)+A_{G}(\rho)\ .
      \label{A_lo_is_A_sh_plus_A_G}
\eeqa
As $A^{\rm (sh)}_{G\otimes G}(\rho)\ge 0$ and $A_{G}(\rho)\ge 0$ then clearly
\beq
   0 \le A^{\rm (sh)}_{G\otimes G}(\rho) \le A^{\rm (lo)}_{G\otimes G}(\rho)\ .
\eeq

The local asymmetry $A_{G\otimes G}^{\rm (lo)}(\rho)$ is the asymmetry of $\rho$ with respect to the local $G$-SSR
which restricts our knowledge of the state to ${\cal G}_{G\otimes G}[\rho]$.  It is clearly related to the total extractable
work $W_{G\otimes G}(\rho)$ that is represented by the state ${\cal G}_{G\otimes G}[\rho]$.  Indeed from
Eqs.~(\ref{W}), (\ref{A_lo}) and (\ref{W_GxG}) we find
\beq
   W(\rho)=W_{G\otimes G}(\rho)+A_{G\otimes G}^{\rm (lo)}(\rho)\ .
   \label{W_is_W_GxG_A_GxG}
\eeq

We can now list the properties of the local asymmetry as
\begin{list}{({\em\roman{enumi}})}{\usecounter{enumi}}
   \item $A_{G\otimes G}^{\rm (lo)}(\rho) \geq 0$~;
   \item $A_{G\otimes G}^{\rm (lo)}(\rho) = 0$ iff
   \beq
     {\cal G}_{G\otimes \{e\}}[\rho] = \rho  \label{A_lo_II_a}
   \eeq
   {\em and}
   \beq
      {\cal G}_{\{e\}\otimes G}[\rho] = \rho  \label{A_lo_II_b}\ ;
   \eeq
   \item $A_{G\otimes G}^{\rm (lo)}(\rho)$ is non-increasing on average under {\em local} $G$-invariant LOCC; and
    \item $A_{G\otimes G}^{\rm (lo)}(\rho)$ is an achievable upper bound on the
    synergy of the extractable work $W_{G\otimes G}$ under the local $G$-SSR restriction.
\end{list}
Once again the first property follows from the properties of the entropy function. The proof of the
third property is a minor modification of the proof of the third property of the shared asymmetry
$A^{\rm (sh)}_{G\otimes G}$.  Similarly, the proof of the fourth property is of the same form as
that of the fourth property of the asymmetry $A_G$.  We leave these proofs for the interested
reader.

The second property can be proved as follows.  We note that the conditions Eqs.~(\ref{A_lo_II_a})
and (\ref{A_lo_II_b}) taken together imply ${\cal G}_{G\otimes G}[\rho]=\rho$ for which $A^{\rm
(lo)}_{G\otimes G}(\rho)=0$ according to \eqr{A_lo} and so the conditions are sufficient. Also the
concavity of the entropy function yields $S({\cal G}_{G\otimes G}[\rho])\ge S({\cal G}_{G\otimes
\{e\}}[\rho])\ge S(\rho)$. Thus $A^{\rm (lo)}_{G\otimes G}(\rho)=0$ implies $S({\cal G}_{G\otimes
\{e\}}[\rho])= S(\rho)$ and hence ${\cal G}_{G\otimes \{e\}}[\rho]= \rho$.  By a similar argument,
$A^{\rm (lo)}_{G\otimes G}(\rho)=0$ implies that ${\cal G}_{\{e\}\otimes G}[\rho]= \rho$. The
conditions are therefore necessary as well.

Either condition \erf{A_lo_II_a} or \erf{A_lo_II_b} is sufficient for ${\cal G}_{G\otimes
G}[\rho]={\cal G}_G[\rho]$ and thus {\em sufficient} for $A^{\rm (sh)}_{G\otimes G}[\rho]=0$.  But
both these conditions are {\em not necessary} for $A^{\rm (sh)}_{G\otimes G}[\rho]=0$.  This means
there is a wider class of states for which $A^{\rm (lo)}_{G\otimes G}[\rho]\ne 0$ than for $A^{\rm
(sh)}_{G\otimes G}[\rho]\ne 0$.

Finally, Eqs.~(\ref{W_GxG}) and (\ref{W_is_W_GxG_A_GxG}) together yield
\beq
   W(\rho^\beta) = W_{G\otimes G-{\rm L}}(\rho^\beta)+E_{G\otimes G}(\rho^\beta)+A^{\rm (lo)}_{G\otimes G}(\rho^\beta)
\eeq
which is consistent with \eqr{comp_E} on recalling \eqr{A_lo_is_A_sh_plus_A_G} and \eqr{comp_W} and
the fact that $A_G(\rho^\beta)=0$ for the globally symmetric state $\rho^\beta$.

\section{Discussion}

In this paper we have quantified the ability of a system to act as a reference system and ameliorate the effect of the
superselection rule $G$-SSR induced by $G$.  Our approach is to express the reference-frame ability of a system in
terms of a physical quantity, namely, in terms of how the system can increase the amount of work that can extracted from
a thermal reservoir. To do this we introduced the quantity $\Syn$ in \erf{DeltaX}, which we call the synergy of two
systems.  The work synergy is the extra amount of work that is extractable using the two systems collectively compared
to the total amount of work extractable using the systems separately. Theorem 2 shows that this quantity is bounded
above by the asymmetry $A_G$ with respect to symmetry group $G$ of each system, a result which elevates the
asymmetry of a system to a resource for overcoming the restrictions of the $G$-SSR. We used the same approach for
bipartite systems where we found (Theorem 3) that the synergy bounds the shared asymmetry $A^{\rm (sh)}_{G\otimes
G}$.

Our results can be arranged in terms of a hierarchy of increasing restrictions, from global $G$-SSR, global and local
$G$-SSR and finally global and local $G$-SSR and LOCC. At each level of restriction we find that the resources
reappear in new forms. For example, under the global $G$-SSR \eqr{comp_W} shows that the unconstrained extractable
work $W$ splits into two new resources of extractable work $W_G$ and asymmetry $A_G$, i.e.
\beq
   W(\rho) = W_G(\rho)+A_G(\rho)
   \label{compW_prime}
\eeq
for arbitrary states $\rho$.  Next under global and local $G$-SSR we find from Eqs.~(\ref{compW_prime}),
(\ref{W_is_W_GxG_A_GxG}) and (\ref{A_lo_is_A_sh_plus_A_G}) that the extractable work $W_G$ further splits
into a more constrained extractable work $W_{G\otimes G}$ and a new asymmetry $A^{\rm (sh)}_{G\otimes G}$, i.e.
\beq
   W_G(\rho) = W_{G\otimes G}(\rho)+A^{\rm (sh)}_{G\otimes G}(\rho)
\eeq
also for arbitrary states $\rho$. Finally under global and local $G$-SSR and LOCC we found
\beq
   W_{G\otimes G}(\rho^\beta) = W_{G\otimes G-{\rm L}}(\rho^\beta)+E_{G\otimes G}(\rho^\beta)
   \label{G_GxG_LOCC}
\eeq
for globally symmetric states $\rho^\beta$. These results are summarized in Table
\ref{table_G_GxG_L}.  A different ordering of the constraints, where LOCC is applied first followed by the global
$G$-SSR and then the local $G$-SSR leads to the results in Table \ref{table_L_G_GxG}.

%

\begin{table}
  \centering
      \begin{tabular}{|c|c|c|}\hline
       Constraints & Resources & State\\
        \hline
         --  & $W$ & $\rho$\\
        $G$ & $W=W_{ G}+A_{ G}$ & $\rho$\\
        $G$ \& $G\otimes G$ & $W_G=W_{G\otimes G}+A^{\rm (sh)}_{G\otimes G}$ & $\rho$\\
        $\left.\begin{array}{c} G, G\otimes G\\ \&\ {\rm L}\end{array}\right\}$
            & $W_{G\otimes G}=W_{G\otimes G-{\rm L}}+E_{G\otimes G}$ & $\rho^\beta$ \\
        \hline
      \end{tabular}
      \caption{Hierarchy of constraints and resources for classes of states where $\rho$ represents an arbitrary state
      and $\rho^\beta$ represents a pure $G$-invariant state.} \label{table_G_GxG_L}
\end{table}

\begin{table}
  \centering
      \begin{tabular}{|c|c|c|}\hline
       Constraints & Resources & State\\
        \hline
         --  & $W$ & $\rho$\\
        ${\rm L}$ & $W=W_{\rm L}+E$ & $\rho$\\
        ${\rm L}\ \&\ G$ & $W_{G}=W_{G-\rm L}+E_G$ & $\rho^\beta$\\
        $\left.\begin{array}{c} {\rm L}, G\ \&\\ G\otimes G \end{array}\right\}$
            & $\begin{array}{l} W_{G-\rm L}+E_G\\
                         \quad\quad=W_{G\otimes G-{\rm L}}+E_{G\otimes G}+A^{\rm (sh)}_{G\otimes G}\end{array}$
                        & $\rho^\beta$ \\
        \hline
      \end{tabular}

      \caption{Hierarchy for a different ordering of the constraints. The equation in the third row
      is obtained directly from the second row using the fact that $W_G(\rho^\beta)=W(\rho^\beta)$,
      $W_{G-{\rm L}}(\rho^\beta)=W_{\rm L}(\rho^\beta)$ and $E_G(\rho^\beta)=E(\rho^\beta)$
      for $G$-invariant states $\rho^\beta$.}\label{table_L_G_GxG}
\end{table}

Our relations Eqs. ({\ref{comp_W}) and (\ref{comp_E}) show the mutually competing nature of the mechanical,
logical and asymmetry resources represented by a state. They are analogous to the particle--wave duality in the following
sense. Asymmetry with respect to the group $G=\{g\}$ can be thought of as a generalized measure of {\it localization}
in that the most asymmetric pure state is transformed into an orthogonal state by the group elements $g$ which is
analogous to moving a particle from one distinct path to another in a which-way experiment. On the other hand,
extractable work under the $G$-SSR measures the invariance of a state to the group action and can be thought of as a
measure of the system's ability to display {\it interference}. Our relation Eq.~(\ref{comp_W}) between asymmetry and
extractable work can then be seen to express a tradeoff between generalized measures of localization and interference.
This connection has been explored elsewhere \cite{JV}.

SSRs are ubiquitous in quantum physics where, for example, spatial orientation is limited by a SU($n$)-SSR and optical
phase is limited by a U(1)-SSR. In the presence of SSRs, quantum states require sufficient asymmetry in their attendant
reference systems in order to be useful.  Moreover, a comparison of the relative efficiencies of classical and quantum
algorithms needs to account for the total amount of resources needed in each case. Quantum reference systems are
clearly a resource that needs to be tallied and so, in this sense, our results pave the way for evaluating the full cost of
resources needed for quantum information processing.  They also open up a new direction of research in the study of
SSRs and reference systems.

We thank M. Plenio and S.D. Bartlett for discussions. This work was supported by the Australian Research Council, the
Queensland State Government and the Leverhulme Trust of the UK.

\appendix
\section*{Appendix}

\subsection{Proof of \eqr{psi}}\label{app_global_symm}

We show that $\ket{\psi^{\mu,\beta}_{m_\mu,m_{\bar{\mu}}}}$ in \eqr{psi} is globally symmetric. From
Eqs.~(\ref{T_reduced}), (\ref{R_beta}) and (\ref{psi}), the scalar nature of $T^\beta$ and the unitarity of the operators
$C^\mu$ and $T^\mu(g)$ we find that
\begin{widetext}
\beqan
      T(g)\ket{\psi^{\mu,\beta}_{m_\mu,m_{\bar{\mu}}}}
      &=&  \bigoplus_{i=1}^{K_{\rm A}}\bigoplus_{j=1}^{K_{\rm B}}
                T^{f_{\rm A}(i)}(g)\otimes T^{f_{\rm
                B}(j)}(g)\ket{\psi^{\mu,\beta}_{m_\mu,m_{\bar{\mu}}}}\nn\\
      &=&  T^{\mu}(g)\otimes T^{\bar{\mu}}(g)\ket{\psi^{\mu,\beta}_{m_\mu,m_{\bar{\mu}}}}\nn\\
      &=&  \left[C^\mu T^\beta(g)[T^{\bar{\mu}}(g)]^* (C^\mu)^{\dagger}\right]\otimes
                  T^{\bar{\mu}}(g)\ket{\psi^{\mu,\beta}_{m_\mu,m_{\bar{\mu}}}}\\
      &=&  \frac{1}{\sqrt{D_\mu}}\sum_{i,j}{C^\mu_{i,j}} C^\mu T^\beta(g)[T^{\bar{\mu}}(g)]^* (C^\mu)^{\dagger}
                \ket{\mu,m_\mu,i}\otimes T^{\bar{\mu}}(g)\ket{\bar{\mu},m_{\bar{\mu}},j}\\
      &=&  \frac{1}{\sqrt{D_\mu}}\sum_{i,j,k,l,n,p}{C^\mu_{i,j}} C^\mu_{n,l} T^\beta(g)
                [T^{\bar{\mu}}_{l,k}(g)]^* (C^\mu_{i,k})^* T^{\bar{\mu}}_{p,j}(g)\ket{\mu,m_\mu,n}\otimes
                \ket{\bar{\mu},m_{\bar{\mu}},p}\\
      &=&  \frac{1}{\sqrt{D_\mu}}\sum_{n,p} C^\mu_{n,l} T^\beta(g)
                    \ket{\mu,m_\mu,n}\otimes \ket{\bar{\mu},m_{\bar{\mu}},p}\\
      &=&  T^\beta(g)\ket{\psi^{\mu,\beta}_{m_\mu,m_{\bar{\mu}}}}\\
      &=&  \lambda^\beta(g)\ket{\psi^{\mu,\beta}_{m_\mu,m_{\bar{\mu}}}}
\eeqan
where  $\lambda^\beta(g)$ is an eigenvalue of unit modulus.  Thus
\beq
    T(g)\left(\uplus\ket{\psi^{\mu,\beta}_{m_\mu,m_{\bar{\mu}}}}\right)T^\dagger(g) =\uplus\ket{\psi^{\mu,\beta}_{m_\mu,m_{\bar{\mu}}}}
\eeq
which completes the proof that $\ket{\psi^{\mu,\beta}_{m_\mu,m_{\bar{\mu}}}}$ is globally symmetric.
\end{widetext}

\subsection{Measurement of charge under the local $G$-SSR}\label{app_charge_meas}

We show here that a local measurement of charge is locally $G$-invariant and thus it is an allowed operation under the
local $G$-SSR.  Let the projection operators onto the flavor and color subsystems be
\beqa
   \openone^{\rm (fl)}_\mu &=& \sum_{m_\mu=1}^{M^\mu}\ket{\mu,m_\mu}\bra{\mu,m_\mu}\\
   \openone^{\rm (co)}_\mu &=& \sum_{i=1}^{D_\mu}\ket{\mu,i}\bra{\mu,i}\ ,
   \label{identity_co}
\eeqa
respectively, where the states $\ket{\mu,i}$ and $\ket{\mu,m_\mu}$ are defined by \eqr{charge_flavour} and $D_\mu$
is the dimension and $M^\mu$ is the multiplicity of the irrep labeled by $\mu$. We note from \eqr{T_reduced} that the
operator $(\openone^{\rm (fl)}_\mu\otimes\openone^{\rm (co)}_\mu)$ ``picks out'' the irrep $T^\mu$ in the following
sense
\beqa
    T(g)(\openone^{\rm (fl)}_\mu\otimes\openone^{\rm (co)}_\mu)
    &=& T^\mu(g)(\openone^{\rm (fl)}_\mu\otimes\openone^{\rm (co)}_\mu)\\
    &=& (\openone^{\rm (fl)}_\mu\otimes\openone^{\rm (co)}_\mu)T^\mu(g)\\
    &=& (\openone^{\rm (fl)}_\mu\otimes\openone^{\rm (co)}_\mu)T(g)
\eeqa
and so $[T(g),\ (\openone^{\rm (fl)}_\mu\otimes\openone^{\rm (co)}_\mu)]=0$. A local measurement by Alice that
projects onto the charge $\mu$ is described by the set of projection operators
\beq
    \Pi_\mu =(\openone^{\rm (fl)}_\mu\otimes\openone^{\rm (co)}_\mu)_{\rm A}\otimes \openone_{\rm B}
\eeq
for $\mu=1,2,\ldots,N_G$ where $N_G$ is the number of irreps of $G$ and subscripts A and B refer to operators
acting on Alice's and Bob's subsystems, respectively. We find that
\beqan
 &&T_{\rm A}(g)\otimes T_{\rm B}(g')
      \Big(\Pi_\mu \rho \Pi_\mu\Big)
      T\dg_{\rm A}(g)\otimes T\dg_{\rm B}(g')\\
   &&=
      \Pi_\mu \left[T^{\phantom{\dagger}}_{\rm A}(g)\otimes T_{\rm B}(g')\right]
      \rho
      \left[T\dg_{\rm A}(g)\otimes T\dg_{\rm B}(g')\right] \Pi_\mu
\eeqan
and so the local projection measurement of charge given by $\{\Pi_\mu: \mu=1,2,\ldots,N_G\}$ is locally
$G$-invariant according to \eqr{local_G_O_AB}.


\begin{thebibliography}{99}
\bibitem{LieGroupNote}
Where we give expressions for discrete groups, the corresponding analysis for compact Lie groups is obtained by
replacing summations with appropriate integrals {\em etc.}
\end{thebibliography}
\end{document}